\documentclass{iopjournal1}
\usepackage[version=4]{mhchem}
\usepackage{graphicx}
\usepackage{subcaption}
\usepackage{wrapfig}

\begin{document}

\title{Toward a Theoretical Roadmap for Organic Memristive Materials}

\author{Salvador Cardona-Serra$^{1,*}$}

\affil{$^{1,*}$Departmento de Química Física, Universitat de València, Valencia, Spain}

\email{salvador.cardona@uv.es}


\begin{abstract}
Neuromorphic computing aspires to overcome the intrinsic inefficiencies of von Neumann architectures by co-locating memory and computation in physical devices that emulate biological neurons and synapses. Memristive materials stand at the core of this paradigm, enabling non-volatile, history-dependent electronic responses. While inorganic oxides currently dominate the field, molecular and polymeric systems can offer untapped advantages in terms of chemical tunability, structural flexibility, low-cost processing, and biocompatibility. However, progress has been hindered by the absence of a theoretical framework able to rationalize how molecular structure translates into memristive function. Here, a multiscale computational perspective is presented, outlining how quantum chemistry and molecular dynamics, among other approaches, can be integrated into a coherent methodology to design next-generation organic memristors. Three mechanisms—ionic migration, redox-driven switching, and conduction interplay in chiral molecules are examined as representative routes toward molecular neuromorphic hardware. The opportunities and challenges associated with each mechanism are discussed, together with a view on how a theoretically guided roadmap can accelerate the emergence of chemically engineered synaptic materials.
\end{abstract}

\section{Introduction}
A computer is composed of two basic and separated elements, the processing, and the memory units. Such a structure has provided an enormous success to CMOS (Complementary metal-oxide semiconductor) electronics. However, for certain leading computer applications, such as pattern recognition and data mining, we face the von Neumann bottleneck, where this separation is inefficient.  The way in which biological neurons work allows to overcome this problem. Thus, there exists a huge effort to build Artificial Neural Networks which mimic the behavior of neurons by using electronic components. For this hardware-based neuromorphic computing, a key microelectronic element is the memristor, which varies non-linearly its resistance according to the previous history of voltage applied to it.\cite{Chua1971}  Memristors are the most suitable components for energy-efficient neurohardware applications as they can combine information processing with memory storage in a single component.\cite{Prezioso2015}  On its original formulation, the memristor aroused from a mathematical reasoning as the functional relation between the electronic parameters charge and flux (representing the time integral of the element’s current and voltage respectively). In terms of real application, the characteristic that defines a memristive material is its ability to vary non-linearly its resistance according to previous history of voltage applied to it over a period of time. Memristors show a distinctive fingerprint characterized by a pinched hysteresis loop in the I-V measurement. This property, called ‘memristivity’ has been proposed to be the foundation of vanguard computer science, from the design of new Resistive RAM (RRAM) that can surpass current Flash Disks and HDD advantages, to the opening of new in-memory computing paradigms. 

The scientific and technological relevance of this idea arises from the envisaged development of memristive materials for new neuromorphic applications. Interest from the European Commission (EC) in this field is well established, as illustrated by the strategic role assigned to neuromorphic technologies within the Future Emerging Technologies (FET) Flagship of the Human Brain Project,\cite{brain2016} one of the three Flagship initiatives under Horizon 2020. This large-scale programme has explicitly promoted the use of multiscale modeling approaches, closely aligned with those adopted in this work, to elucidate artificial neuronal behavior. In its roadmap, the EC has identified the memristor as the most promising building block for next-generation neuromorphic systems, noting that its successful implementation could provide Europe with a decisive technological advantage.

Substantial progress in brain-inspired computing has also been achieved by leading technology companies such as IBM, Intel, and HP. Although most commercial computing hardware still relies on conventional CMOS transistor-based architectures, recent neuromorphic platforms, such as SpiNNaker,\cite{Spinnaker2020} BrainScaleS,\cite{brainscales2022} and IBM’s TrueNorth\cite{truenorth2015} chip, have gained prominence for large-scale data analysis by using spiking neuron paradigms to emulate biological information processing. These systems have demonstrated several advantages, but their transistor-based neurons ultimately remain constrained by intrinsic CMOS limitations.

Achieving a more faithful emulation of the brain will require not only new computational architectures but also dedicated materials capable of reproducing key neuronal functions. This is precisely where memristors, and particularly molecular memristors, are expected to play a transformative role. Their theoretical limits in terms of miniaturization, energy efficiency, and performance far exceed those attainable with transistor-based technologies. As an illustration, solid-state memristors have already demonstrated neuron-like elements as small as 2 nm,\cite{Pi2019} peak programming powers of 0.23~$\mu$W, and storage densities surpassing 4.5 terabits per square inch.

The rapid expansion of neuromorphic technologies has also highlighted a crucial scientific gap. Inorganic memristors have reached impressive performance, but their chemical rigidity, limited tunability and difficulties in ensuring reproducible nanoscale behavior restrict their long-term potential. Organic and molecular systems emerge as an appealing alternative, yet the same features that make them attractive, structural diversity, soft-matter dynamics and chemical adaptability, also complicate the establishment of clear structure–function relationships.\cite{vandeBurgt2018, Tuchman2020, Hoch2025} As a result, the field still lacks a coherent theoretical vision capable of guiding the design of new materials with genuine predictive power.

In this Perspective, molecular memristors are placed within this broader technological context with the aim of clarifying both their opportunities and their challenges.\cite{lee_2025} Progress in this area will require an integrated, multiscale approach that connects electronic structure at the quantum level with the emergent behavior observed in operating devices. The purpose here, is to highlight the recurring principles that underlie memristive function across different chemical platforms and to identify the conceptual obstacles that must be addressed before organic neuromorphic components can transition from promising prototypes to reliable technologies. The following sections reflect this vision: we first outline the capabilities and limitations of inorganic systems, then review the main mechanisms that enable memristivity in organic and molecular materials, and finally propose a theoretical strategy that can serve as a foundation for the rational design of future organic neuromorphic devices.

\subsection{Inorganic candidates for memristive materials}
A large variety of memristive materials are available,  but most of the current research is focused on the use of extended inorganic materials.\cite{Strukov2008, Cho2016, Lee2011, Chanthbouala2012a} The first memristive material was proposal as late as in 2008 in HP laboratories.\cite{Strukov2008} This device was strongly grounded in the previous knowledge of resistive switching materials. It was experimentally prepared by sandwiching a thin layer of titanium oxide (3-30nm of \ce{TiO2}) between two metallic Ti electrodes. This insulator oxide was doped with oxygen vacancies (n-type) thus acquiring semiconductor behavior in the reduced sublayer \ce{TiO2-x}. The presence of these vacancies, which can be thermally or electrically displaced, was enough to produce structural modifications leading to notable resistive changes (on/off ratios of more than three orders of magnitude) with fast writing times ($\approx$ ns) and retention times in years-scale. It is important to focus on the enormous electric fields that act on these very thin layers ($<1$MV/cm), which elicit an effective reduction in the activation barrier for ion migration inside the solid, and therefore, a strong non-linearity in the conductance.
Since this discovery, new attempts have been made to improve the reproducibility and retention time of these memristive materials.\cite{carlos2021, Khan2021, Kim_2011} Another particularly successful material is based on the so-called ‘Phase Change Memories’ (PCM).\cite{Koelmans2015, LeGallo2018} These materials were initially suggested as new-generation non-volatile memories which could vary their electrical resistivity by changing its structure from a crystalline to an amorphous phase.\cite{Sebastian2014}  Typically, when a strong enough voltage pulse (RESET pulse) is applied, the crystal locally melts due to Joule heating. Thus, as the voltage is stopped abruptly, the molten material forms an amorphous glassy phase which blocks the conduction in the junction. The size of this phase depends on the amplitude and the width of such pulse. Finally, to recover the initial state, another voltage pulse (SET pulse) is applied ensuring that the reached temperature is not so high as to melt the material but sufficient to crystallize the amorphous phase. 
A third type of memristive materials uses nanoscale spintronic oscillators, a totally different approach, where magnetism and electronics interplay for building the neuronal units.\cite{Torrejon2017}  This route is based on the concept of spin-valve magnetoresistance, where the total resistance of the Magnetic Tunnel Junction (MTJ) depends on the relative orientation of a soft-variable ferromagnet with respect to a hard-fixed ferromagnet.\cite{Chanthbouala2012b}  Thus, the current passing through the junction generates a torque on the magnetization of the soft-ferromagnet which leads to a spin precession with frequencies varying from 100 MHz to tens of GHz. These spin precessions can be converted to voltage oscillations through magnetoresistance.\cite{Romera2018}  Due to their response to thermal noise, MTJ have been presented as non-volatile magnetic memories that can contribute with a certain stochasticity in the resistance transition.\cite{Liu2024, Oberbauer2025} This property has been recently exploited for random number generation and other noise-based computing applications. \cite{Mizrahi2018} 
Despite the recent success of this type of materials, they are inherently limited by a low chemical variability and functionalization, by difficulties in nanostructuration, by poor reproducibility because of large parameter dependences and by limited cycling endurance.\cite{Mohammad2016, Wang2018} In contrary, molecular materials are potential memristive materials which excel in some of these aspects. First, they allow using the versatility of molecular chemistry to design and tailor the exhibited properties.\cite{Duan2012} Second, they permit to reduce the size and the energetic requirements.\cite{jeong2017} Third, they improve the processability by means of low-cost solution fabrication processes (as surface deposition for example).\cite{chen2013,garish2024} Among these, there are other useful properties exhibited by organic materials such as transparency, flexibility, environment friendly and, in several cases, biocompatibility.\cite{Kim2022, Krauhausen2021} In order to improve these materials and their properties, it is crucial to rely on theoretical studies, however building a  theoretical design roadmap for molecule-based memristors has been elusive yet.\cite{Zeng2014,Li2013, Gkoupidenis2017, Al-Bustami2018,Miyamachi2012}  A few existing theoretical studies are based on HOMO/LUMO level  determination,\cite{Goswami2017} molecular dynamics simulations,\cite{Savelev2011}  or atomistic molecular relaxations\cite{Agapito2009}  and so far have been very efficient to understand the relationship between chemical structure and memristive properties. In the last years, material scientists have claimed that the guidance of an adequate theoretical approach, is mandatory to understand and build robust and efficient memristors.\cite{Sun2019,Lanza2019}

 There are several aspects where theoretical chemistry can contribute into the restructuration of the current way of computing. First, the development of an accurate multiscale methodology to understand molecular memristors will permit the rationalization of the mechanisms up to atomic scale. This initial theoretical issue by itself is an outstanding research concern that has been mostly unexplored up to now. This step will be critical in order to optimize the desired properties thus facilitating the obtention of a material fit for the market. A further advantage is that this methodology could be easily extended to other molecular properties with technological interest, such as spintronics or optoelectronics, so every methodological improvement developed will be useful to other scientists in these fields. An additional step requires a major effort in understanding deeply the effect of an external electric field on a molecular material. Understanding time-reversal symmetry in, i.e. molecular ionic liquids, is crucial to achieve longer retention times (memory effect) and to tune the ON/OFF ratio. In addition to these, being able to tune at will the Fermi energy of a certain metallic electrode by chemical decoration is essential to design efficient electron/hole injector layers. These injection layers represent interfacial materials which facilitate the current to flow from the metallic electrode to a molecular active layer. Concerning redox-active materials, the most crucial issue is to relate the effect of multiple substituents to the presence of various electronic states where the material is stabilized via the application of an external bias. Here, there exist two points that can shed light on the transition mechanism and remain mostly unexplained up to now. The first consists in understanding how the physical crystalline reorganization (in terms of ligands and counterions) occurs when the bias is high enough to force the electronic transition. The second one is associated with the effective coupling between vibrations and electronic transitions and how the former may mediate the electronic excitation. This process is of critical importance in the memristive properties, as Joule heating is strong enough to vibrationally excite the molecular material and intervene in the redox behavior. Finally, achieving memristivity via magnetic interactions in molecules will suppose a severe reduction in terms of size and energy requirements of the active material. Using the spin’s degree of freedom instead of the electron’s charge has been a revolutionary proposal that is currently used in both spintronics and quantum technologies among others. Extending that scheme to memristive materials seems to be the natural road to obtain organized nanomaterials with magneto-electronic properties on a functional device. 

\section{Organic Memristive Materials}

\begin{figure}[ht]
\begin{subfigure}{0.3\textwidth}
\includegraphics[width=0.9\linewidth]{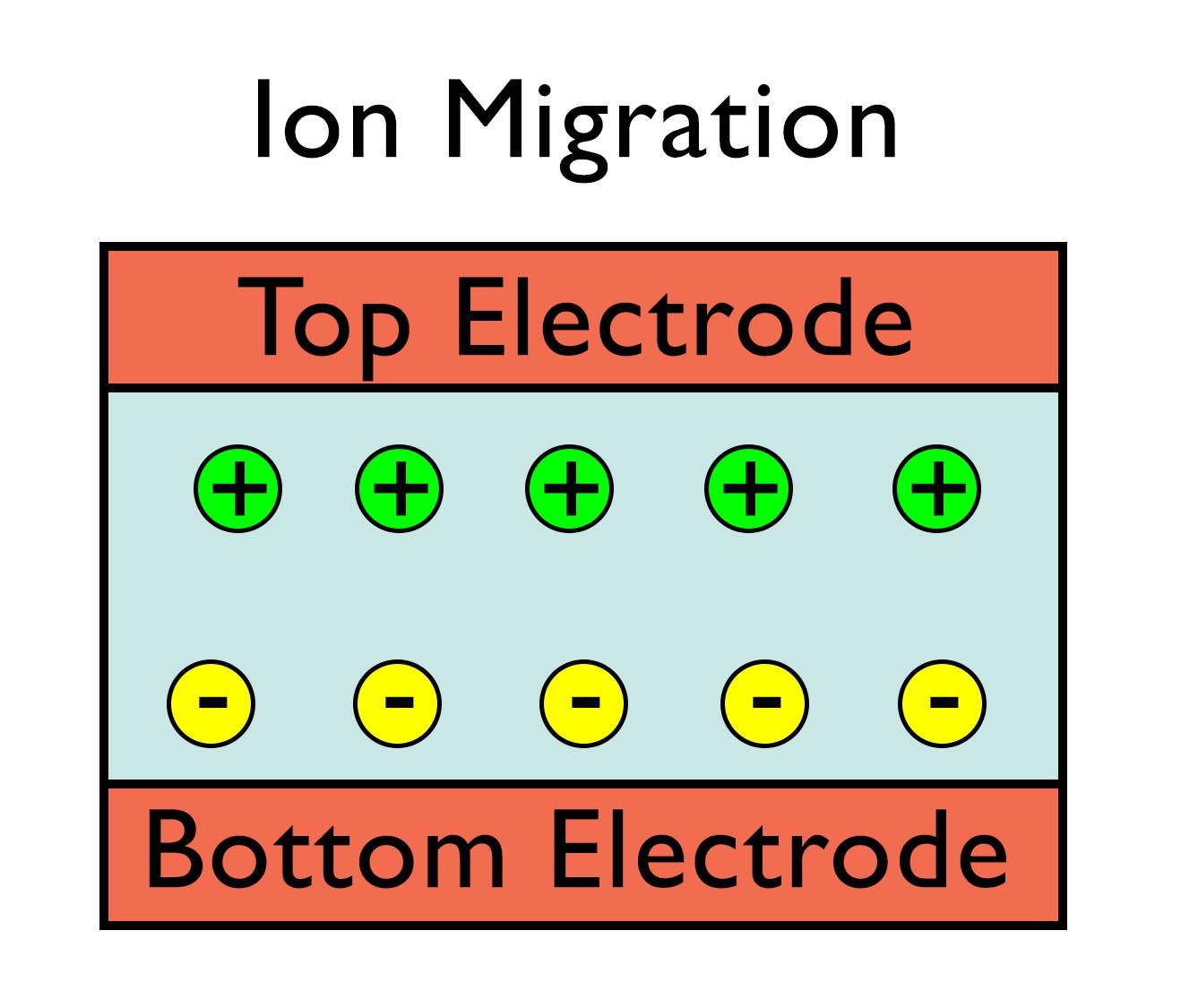} 
\label{fig:a}
\end{subfigure}
\begin{subfigure}{0.3\textwidth}
\includegraphics[width=0.9\linewidth]{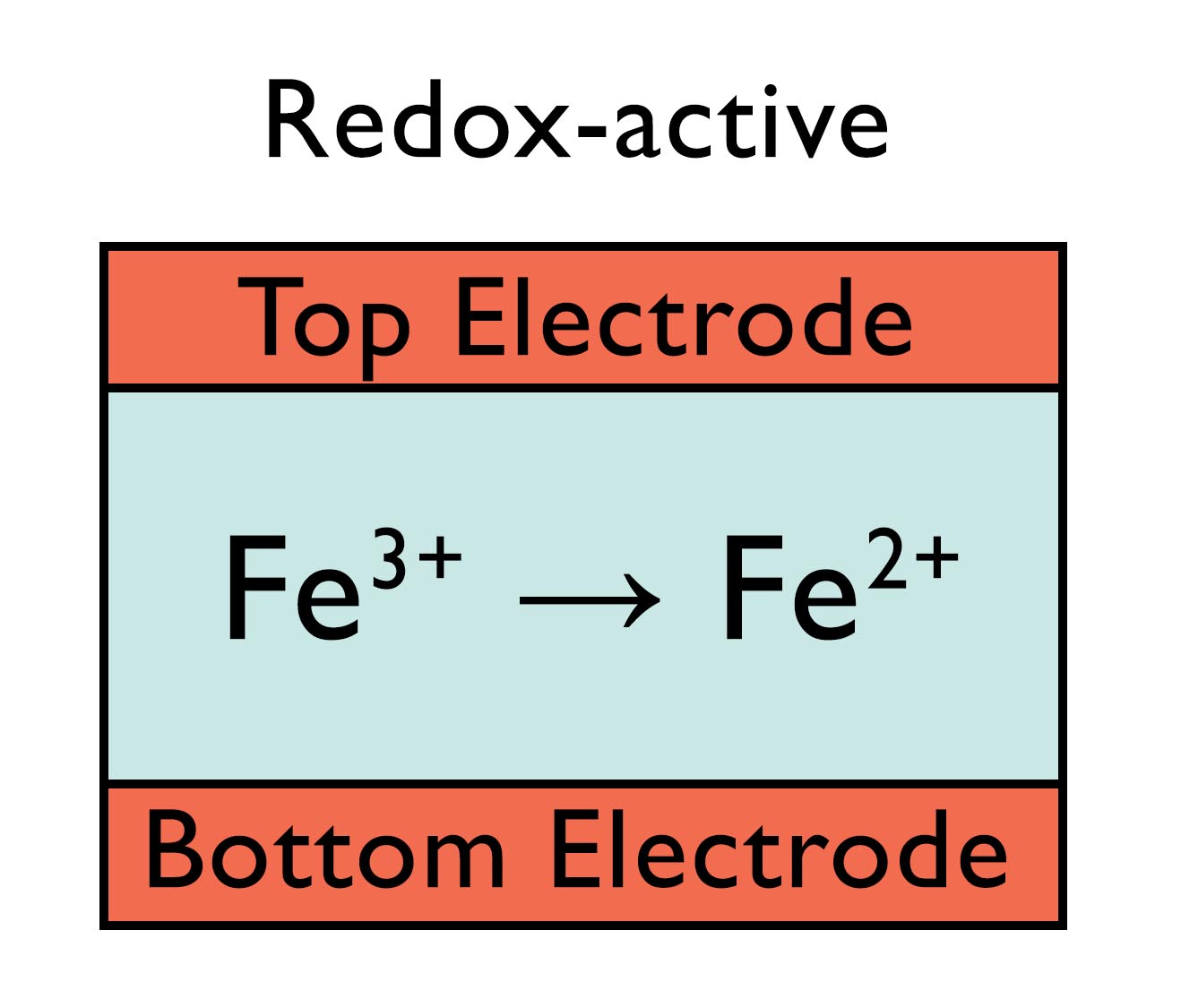}
\label{fig:b}
\end{subfigure}
\begin{subfigure}{0.3\textwidth}
\includegraphics[width=0.9\linewidth]{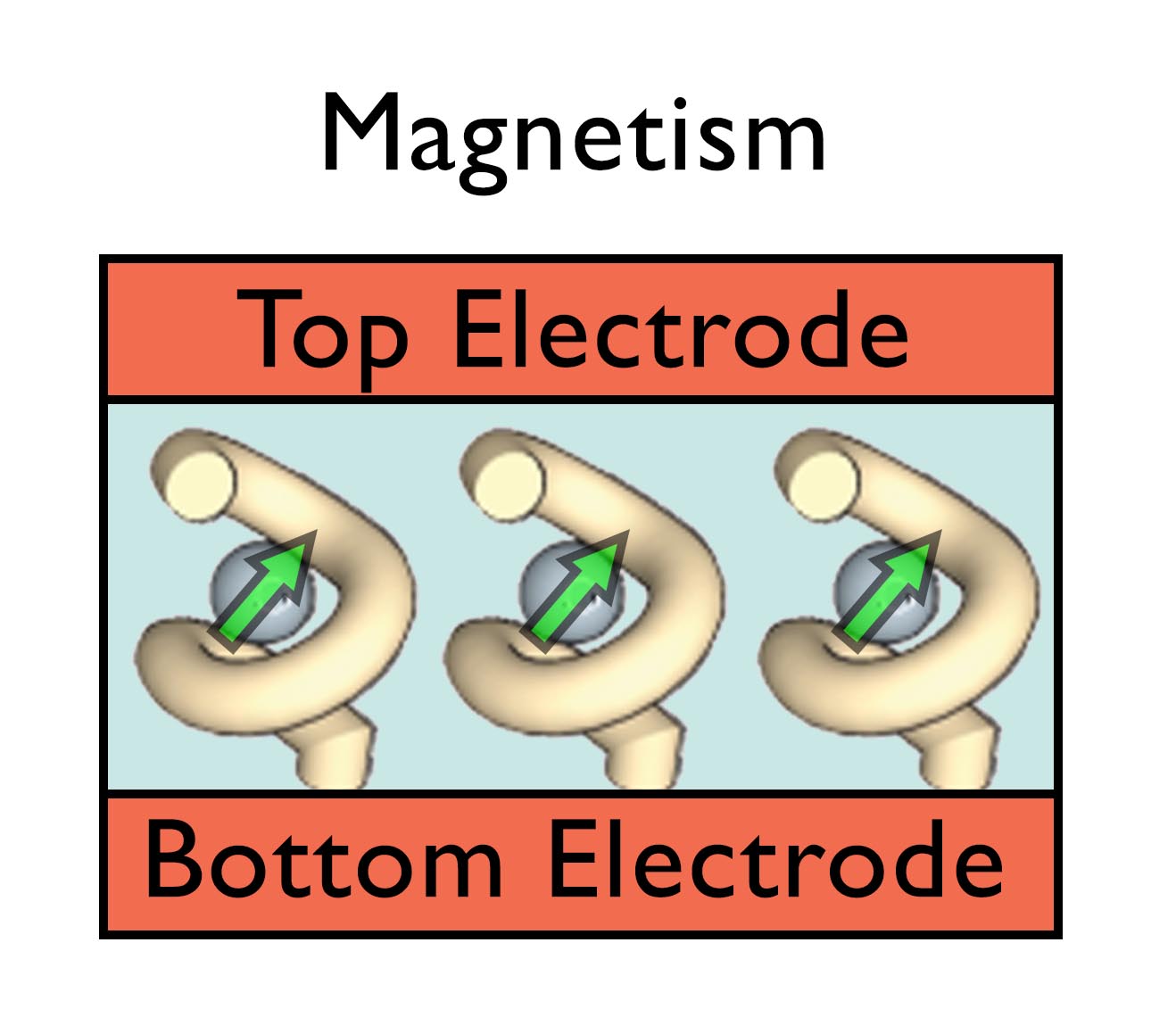}
\label{fig:c}
\end{subfigure}
\caption{Schematic depiction of the  organic materials described in this work as proposals for achieving high-efficient memritive properties}
\label{fig:image2}
\end{figure}

Organic memristive systems offer an attractive alternative to inorganic platforms due to their chemical tunability, low-cost processing, mechanical flexibility and the possibility of encoding switching mechanisms directly into molecular structure.\cite{Williams2024} However, unlike metal oxides or phase-change materials, where a small number of dominant physical processes govern device performance, organic systems exhibit multiple competing pathways that can all give rise to history-dependent conduction. This richness is an opportunity for design, but also a source of complexity that has hindered the establishment of general structure–function relationships. In what follows, three representative mechanisms, ionic migration, redox-based switching and magnetism–chirality interplay, are discussed not only in terms of their operating principles but also through a critical assessment of their current strengths and limitations. (see figure \ref{fig:image2})

Ionic migration constitutes one of the most experimentally established routes to organic memristive behavior, particularly in soft-matter systems where mobile ions coexist with electronically conducting matrices. In these materials, the application of an electric field induces a spatial redistribution of ions, which modulates the local electrostatic landscape and, in turn, the injection barriers at the electrode interfaces.\cite{VanReenen2010, Matyba2009} (see figure \ref{fig:image2} This produces characteristic time-dependent current responses and pinched hysteresis loops that are fully consistent with memristive dynamics. A notable strength of this approach is its compatibility with low-temperature fabrication, solution processing and large-area substrates, making it a natural candidate for scalable neuromorphic hardware.

In the absence of an externally applied voltage or current, ions within the polymer matrix remain uniformly distributed across the device. Once a voltage stimulus is imposed, the resulting electric field induces slight displacements of cations and anions, reshaping the internal potential landscape and giving rise to a charged double layer near the electrodes. As a consequence, the device’s effective resistance becomes dependent on the duration, magnitude, and polarity of the applied bias, endowing the system with a time-dependent internal state governed by the current stimulus.

\begin{figure}[ht]
\begin{subfigure}{0.7\textwidth}
\includegraphics[width=1.0 \linewidth]{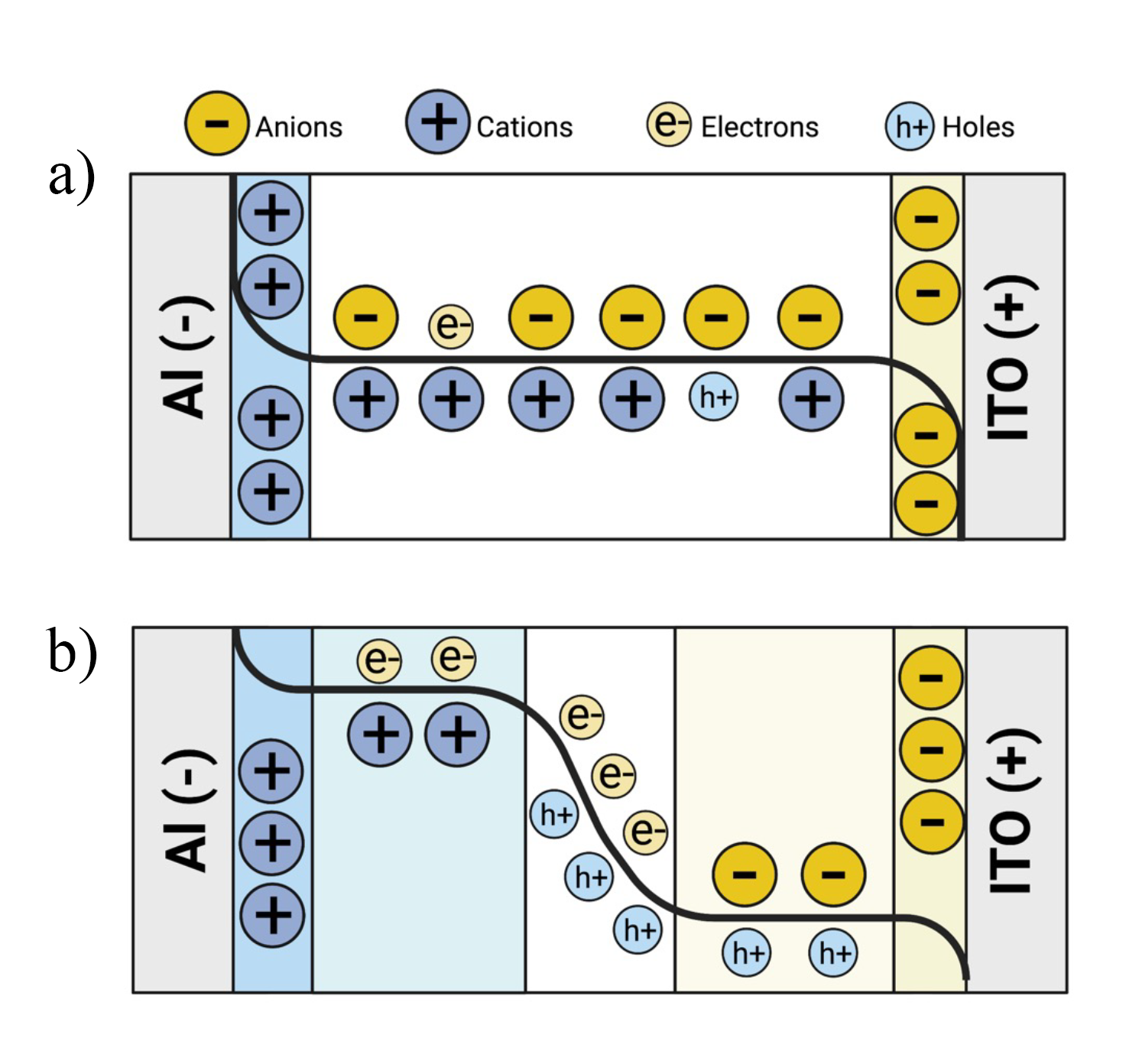}
\label{f}
\end{subfigure}
\caption{Schematic 1D representation of ion migration mechanisms. (a) Electrodynamical model (injection-limited regime, no effective polymer doping exists) (b), Electrochemical model (ohmic injection regime, the polymer conductivity is modified by electronic doping). Reproduced from Ref. \cite{PradoSocorro2022}}
\label{fig:imageafaba}
\end{figure}

Two predominant frameworks have been proposed to describe the motion of ionic species and its implications for charge injection, each supported by experimental evidence yet presenting its own limitations. In the Electrodynamic (ED) model,\cite{DeMello2002} an injection-limited regime is assumed in which most of the potential drop occurs near the electrodes, leaving the electric field nearly negligible in the central portion of the device. The formation of the interfacial ionic double layer locally bends the polymer’s valence and conduction bands, effectively lowering the Schottky barrier for electron and hole injection. Conversely, the Electrochemical Doping (ECD) model describes a scenario where, although ions still accumulate near the contacts and the voltage drop is reduced at the leads, charge injection is ohmic.\cite{pei1995} Under these conditions, electronic doping can proceed, requiring counter-migration of anions and cations to preserve charge neutrality. This redistribution suppresses the electric field throughout most of the structure, except at the central region where recombination takes place (see figure \ref{fig:imageafaba}). After considerable debate, it was eventually clarified in 2010 that the two descriptions correspond to distinct operational regimes determined by the nature of the injection process.\cite{VanReenen2010}

A central aspect of this mechanism is the decoupling of ionic and electronic mobilities, which allows ionic motion to act as the slow internal state variable of the device. By tailoring the chemical nature of the ionic species and the polymer matrix, the timescale of the ionic response can be finely tuned, enabling behaviors that mimic biological synapses such as paired-pulse facilitation, short-term plasticity or adaptation. Systematic modification of the cation and anion, adding ionic liquids, variations in alkyl chain length and the introduction of $\pi$-stacking substituents all strongly influence aggregation behavior, ion-pair lifetimes and barrier-lowering efficiency\cite{vanReenen2013, lenes2011}. This tunability represents one of the main advantages of organic systems over their inorganic counterparts, where the available defect chemistry is typically more rigid.

\begin{figure}[ht]
\begin{subfigure}{1.0\textwidth}
\includegraphics[width=1\linewidth]{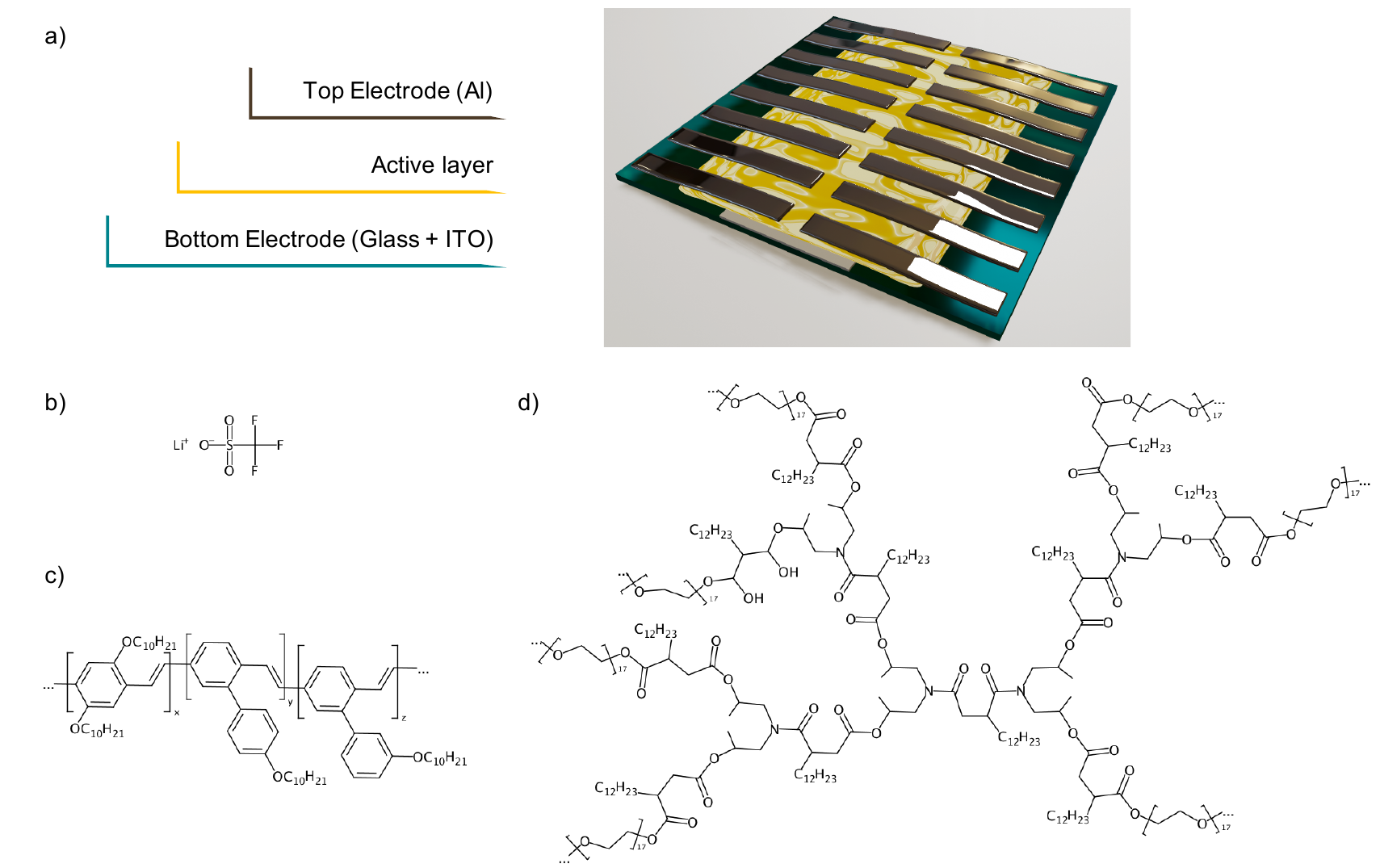} 
\label{e}
\end{subfigure}
\caption{ (a) Schematic of a single synaptic device, where the memristive material shows homogeneous presence of salt ions and composition of the memristive layer: (b) lithium triflate (LiCF3SO3), (c) Super Yellow (semiconductive polymer), and (d) $Hybrane^{\copyright}$ DEO750 8500 (ion-transport polymer).  Reproduced from Ref. \cite{PradoSocorro2022}}
\label{fig:imageafaa}
\end{figure}

An important development in this area has been the proposal by Prado-Socorro et al.,\cite{PradoSocorro2022} who demonstrated that polymer-based materials inspired by light-emitting electrochemical cells (LECs) constitute a particularly promising platform for ionic memristors. LECs are known for their mixed ionic–electronic conductivity and the spontaneous formation of dynamic p–i–n junctions upon biasing. Prado-Socorro and co-workers leveraged these properties to show that similar polymeric blends, typically combining a conjugated polymer with an ionic liquid or a large organic salt, can function as memristive elements when operated under neuromorphic conditions (see figure \ref{fig:imageafaa}). This work highlighted that the ionic accumulation layer formed near the electrode interface in LEC-like systems naturally gives rise to voltage-history-dependent conductance, while the soft reconfiguration of the polymer matrix allows repeated cycling without structural damage. A crucial insight from these studies is that the self-organizing electrochemical junction, long viewed as a feature specific to LEC operation, is in fact a powerful mechanism for achieving controllable ionic migration in memristive devices.\cite{VanReenen2010, meier2012J} This discovery opened a route to designing memristors that exploit the same interplay between mobile ions and electronic carriers that has made LECs attractive for low-voltage optoelectronics.\cite{tordera2012, meier2014}

Despite their promise, ionic-migration memristors face several challenges that limit their immediate applicability in neuromorphic hardware. Retention times remain strongly influenced by spontaneous back-diffusion of ions once the field is removed, and achieving long-term non-volatility requires either immobilizing ions at the interfaces or engineering energy landscapes that inhibit relaxation. The inherently distributed nature of ionic motion also introduces device-to-device variability, as subtle changes in film morphology, polymer crystallinity or residual solvent content can significantly alter transport pathways.\cite{Liu2023, Song2022} Furthermore, while ionic migration is excellent for short-term plasticity, reproducing long-term synaptic behavior with high endurance remains difficult due to degradation under repeated ionic accumulation and release.

The complexity of the ionic dynamics also complicates compact modelling. Unlike materials governed by a single dominant mechanism, ionic systems typically exhibit a broad hierarchy of relaxation times, arising from local rearrangements, cluster formation, segmental polymer motion and long-range ionic drift. Capturing these multiscale processes in a predictive model suitable for circuit-level simulation represents an open challenge. Nonetheless, the chemical flexibility, low cost and rich dynamical behavior of these materials make ionic-migration-based systems one of the most versatile families of organic memristors, particularly for applications requiring tunable temporal responses, short-term plasticity or dynamic learning rules.

\begin{figure}[ht]
 \centering
       \includegraphics[width=0.8\textwidth]{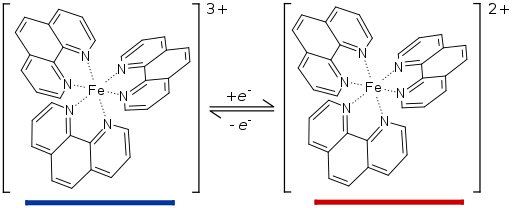}
\caption{Redox-activity of \ce{Fe^{3+/2+}(phen)_3} complex}
\label{figqq}
\end{figure}

Redox-active molecular systems constitute another of the most conceptually appealing routes to organic memristivity because they offer discrete, quantized electronic states that can be addressed electrically.\cite{Goswami2021} In these systems, the application of a bias induces a reversible change in oxidation state of the molecular unit—or an ensemble of units—giving rise to well-defined conductance transitions. (see figure \ref{figqq}) Unlike ionic migration, where the internal state is a continuous variable linked to ion displacement, redox switching can in principle produce digital or multilevel states that are attractive for neuromorphic architectures requiring stable synaptic weights.\cite{Goswami2017, goswami2020b, mahapatra2025}

At the molecular scale, the redox response is governed by a delicate interplay between the intrinsic electronic structure of the active centre and its local environment, which includes ligand conformations, counterion coordination, and intermolecular packing.\cite{Goswami2020} Transition-metal complexes, metallocenes, organometallic sandwich compounds, and $\pi$-conjugated organic molecules with stable radical states have all been shown to exhibit memristive signatures when incorporated into nanoscale junctions. A major advantage is the chemical precision with which redox potentials, reorganization energies, and coupling to electrodes can be tuned.\cite{goswami2020c} This precision enables, at least in principle, the rational design of synaptic behaviors such as multilevel storage, asymmetric weight update, and controlled volatility.

Despite these strengths, redox-based memristors face several challenges related to the coupling between electronic transitions and structural relaxation. Changes in oxidation state are often accompanied by geometric rearrangements, ranging from subtle ligand-field distortions to large counterion reorganizations. These processes can introduce sizeable hysteresis and provide the necessary memory effect, but they also increase switching barriers and reduce endurance. In thin films or polymeric matrices, such reorganizations propagate through the surrounding environment, generating slow relaxation modes that broaden the distribution of switching voltages and compromise reproducibility.

Another persistent difficulty is the heterogeneity of the active volume. While single-molecule devices typically exhibit clean, quantized switching between well-defined redox states, macroscopic films composed of the same molecules frequently show dispersed responses. Variations in packing, grain boundaries, microphase separation and local dielectric environments all influence redox energetics and electron-transfer kinetics. This leads to device-to-device variability and challenges the scalability of redox systems for dense neuromorphic arrays. Achieving uniformity requires precise control over film deposition, solvent evaporation dynamics and the incorporation of counterions with predictable mobility.\cite{rath2023}

Thermal effects further complicate device behavior. Redox transitions in confined organic junctions can be influenced by Joule heating, which may activate unintended vibronic pathways or accelerate degradation reactions. These effects are especially pronounced in high-resistance devices, where local temperatures can rise significantly during switching events. Ensuring long-term chemical stability under cycling thus demands judicious choice of ligands, incorporation of steric protection around the redox centre, and careful control of electrode work functions to minimize overpotential.

A crucial and often underappreciated aspect of redox-based memristors is the role of the ionic environment. Although the active component is electronic in nature, counterions are essential to maintain charge neutrality during oxidation–reduction cycles. The mobility and coordination of these counterions influence both the switching speed and retention time. Systems with loosely bound counterions tend to switch quickly but suffer from volatility, whereas more strongly coordinating counterions enhance retention but slow down switching kinetics. Finding the right balance between these regimes remains a key design challenge and suggests that hybrid redox–ionic mechanisms may offer a promising compromise.

In recent years, coordination complexes designed to minimize structural reorganization have gained attention. Ligands that enforce rigid geometries, low-spin configurations or symmetric charge distribution can suppress large-scale relaxation, enhancing both endurance and reproducibility. Similarly, redox-active polymers and metal–organic frameworks offer pathways to extend molecular-level control to mesoscale assemblies that retain structural order and chemical uniformity.\cite{Torres-Cavanillas2021} These materials may bridge the gap between the precision of single-molecule redox switching and the scalability of solid-state devices.

Overall, redox-based memristors provide a powerful conceptual framework for achieving energy-efficient non-volatile switching, with the added benefit of molecular-level tunability. Yet fully leveraging their potential requires addressing the intrinsic coupling between redox transitions and structural dynamics, improving film-level uniformity, and developing models capable of capturing both fast electronic transitions and slower environmental relaxation processes. As these issues are resolved, redox-active systems may emerge as key building blocks for neuromorphic architectures requiring robust multilevel states and long-term information retention.

A third, and more recently explored, mechanism arises from the interplay between molecular chirality, spin polarization and magnetic moments.
The chiral-induced spin selectivity (CISS) phenomenon has been firmly established across a wide range of biomolecular systems, spanning modified amino acids such as L- and D-stearoyl lysine, short helical peptides like polyalanine, and more complex architectures including bacteriorhodopsin and double-stranded DNA.\cite{naaman2012} Remarkably, however, these demonstrations have almost always relied on diamagnetic scaffolds. In parallel, molecular spintronics studies have revealed that even achiral, non-biological molecules can support spin-dependent transport when a single paramagnetic center is incorporated.\cite{Bogani2008}

This convergence motivates exploring chiral polypeptides containing paramagnetic ions as hybrid spintronic elements.\cite{Torres-Cavanillas2020} In such constructs, the metal center can function analogously to a soft magnetic component, while the intrinsic molecular chirality—through the CISS mechanism—acts as an effective hard magnetic element. Their interplay naturally enables memristive behavior in which the device conductance is conditioned by the past voltage history. The chemical complexity required to produce chiral, multinuclear coordination complexes would be formidable in traditional synthetic chemistry; yet biomolecules offer an alternative route through self-assembly into intricate architectures, including polynuclear, magnetically active, and chiral peptide frameworks. This raises the prospect of employing biomolecular platforms as multistate spintronic memristive units.\cite{Cardona-Serra2021}

\begin{figure}[h]
\begin{subfigure}{0.5\textwidth}
\includegraphics[width=0.8\linewidth]{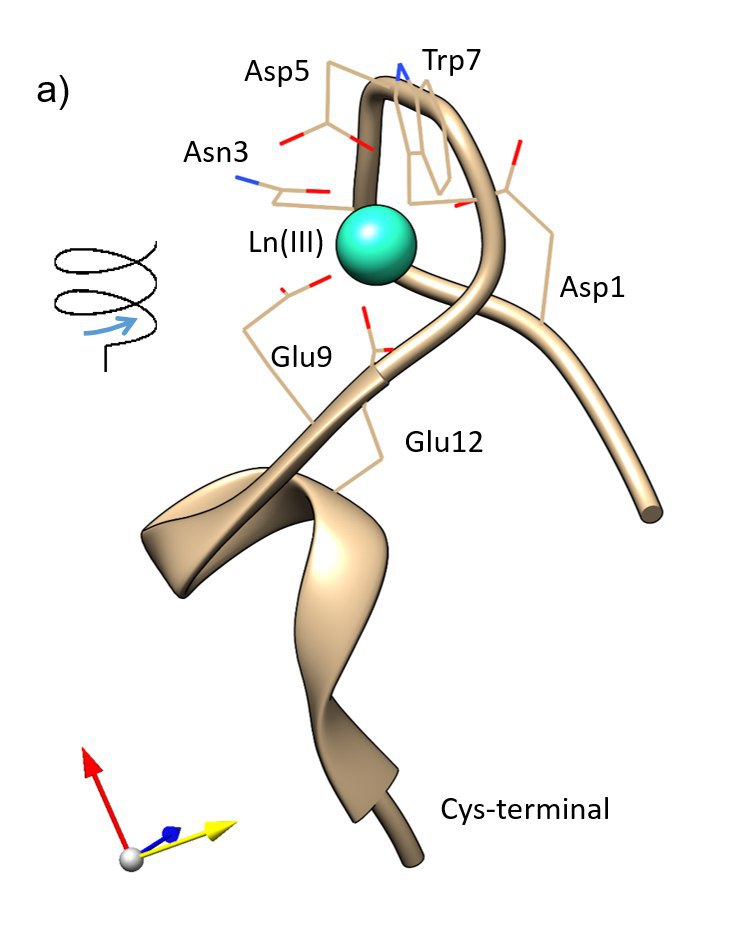} 
\label{c}
\end{subfigure}
\begin{subfigure}{0.5\textwidth}
\includegraphics[width=0.8\linewidth]{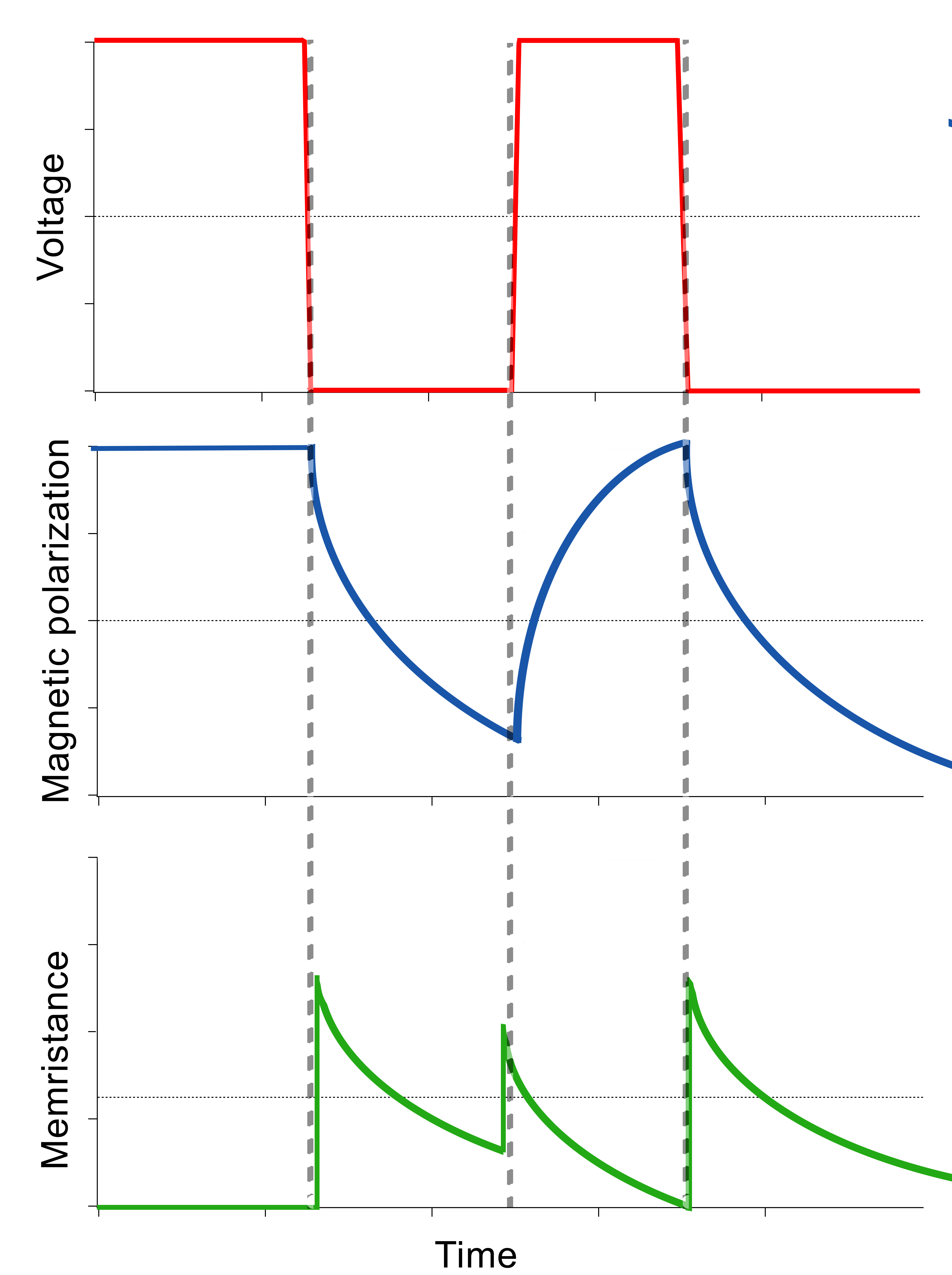}
\label{d}
\end{subfigure}
\caption{ (left) Structure of the LBTC peptide complexed with \ce{Tb^{3+}} (green sphere). In this LBT variant the two C-terminal aminoacids have been replaced with a cysteine residue (right) Schematic functioning of a spintronic multistate memristive device based on the CISS effect in a paramagnetic molecular material. Let us start the experiment with the voltage in the ON position for an unspecified but
long enough time to guarantee that the magnetic polarization of the metal
ions is maximal due to the interaction with the CISS-polarized current, and
the resistance is correspondingly minimal. A sudden switch in the voltage
to the OFF position, e.g. a change in sign has two consequences: it starts a
relatively slow process of relaxation of the magnetic polarization, and it
instantly increases the memristance to its maximum value. This sudden
jump in the memristance is due to the ‘wrong’ polarization of the
paramagnetic ions, which now opposes the spin polarization of the CISS
current. As the magnetic polarization evolves towards its new equilibrium
situation, the memristance decreases too, accessing a continuum of
values, i.e. behaving as a multistate memristance. Depending on the timing
of the subsequent changes in the voltage between the ON and OFF values,
the time evolution of the magnetic polarization of the system and of the
memristance of the device will present different shapes; a fine timecontrol of the voltage allows achieving any desired memristance between
the ON and OFF limit values. Reproduced from Ref. \cite{Cardona-Serra2021} with permission from the PCCP Owner Societies.}
\label{fig:imageaaa}
\end{figure}

Within this broad landscape, virtually any metalloprotein, spin-tagged peptide, or even DNA-origami-based construct could act as a candidate memristive element. Our focus here is on a specific class of systems where the spin of a coordinated lanthanide ion interacts with a spin-polarized current induced by the CISS effect.\cite{Torres-Cavanillas2020} Conceptually, a helical peptide injects spin-polarized carriers, which then interact via magnetic exchange with a localized lanthanide moment. The interplay becomes frequency-dependent: at long times the ion’s average magnetic polarization simply tracks the current direction, but at sufficiently high operating frequencies ($\approx$ GHz regime), the polarization instead reflects the device’s voltage history. (see figure \ref{fig:imageaaa}) Because conductivity depends sensitively on the instantaneous sign of the metal-ion polarization, the overall conductance becomes a history-dependent quantity—a molecular-scale manifestation of memristive behavior.

A recent work has demonstrated this concept using lanthanide-binding tags (LBTs), short helical peptides (15–17 residues) optimized for selective Ln$^{3+}$ coordination and originally developed to provide luminescent tags in proteins. Owing to their genetic encodability, LBTs can be introduced at precise sites in folded proteins or used as isolated peptide ligands. Building on their favorable coordination and spectroscopic properties, we recently reported spin-filtering in solid-state devices assembled from monolayers of LBT peptides complexed with Tb$^{3+}$.\cite{Torres-Cavanillas2020} The contribution of the paramagnetic ion was confirmed through cyclic voltammetry, electrochemical impedance spectroscopy, and local transport measurements using liquid-metal contacts. Spin polarization values approaching 70\% were obtained, comparable to the strongest CISS-based spin filters reported so far, with the advantage that the two sources of spin selectivity—chirality and paramagnetism—can be tuned independently.

From a design perspective, magnetic peptides offer features particularly desirable for neuromorphic-style hardware. The switching time can be adjusted by modifying the ligand field around the lanthanide, which strongly influences its spin relaxation dynamics. The number of accessible states can also be expanded by employing multinuclear designs or enhancing exchange interactions, thereby improving synaptic-like behavior while retaining the low-power advantages inherent to spin-based operation. Although redox-based and ion-migration memristors often achieve large ON/OFF ratios, such limitations in magnetic systems may be mitigated through the incorporation of several exchange-coupled paramagnetic ions.

A longer-term vision is to engineer metallopeptide assemblies containing multiple paramagnetic centers with controllable, site-specific spin dynamics.\cite{Rosaleny2018} Distinct exchange strengths or relaxation times at different sites would naturally yield a spectrum of switching behaviors within a single molecule. By tuning the externally applied voltage pattern, different combinations of high- and low-resistance states could be selectively stabilized, producing rich, multistate response curves. Importantly, this does not require multiple metal species: altering coordination geometry alone by following strategies already established in single-ion magnet research is sufficient to modulate both magnetic anisotropy and relaxation pathways.

Modeling this phenomenon requires a sophisticated combination of electronic-structure calculations including spin–orbit coupling, transport theory capturing spin-dependent hybridization at metal–molecule interfaces, and dynamical magnetic equations evolved under time-varying fields. These materials represent a particularly exciting frontier because they merge organic chemistry, spintronics and neuromorphic engineering within a single molecular entity.

Together, these three mechanisms illustrate the diversity and richness of pathways through which organic materials can achieve memristive function. They also highlight the pressing need for unified theoretical descriptors that allow meaningful comparison and systematic design. Parameters such as ion mobility, redox-level density, magnetic anisotropy, interfacial work-function modification, vibrational coupling strengths, and electronic transmission spectra must be integrated into a common language capable of predicting synaptic metrics such as switching energy, retention time, volatility, conductance linearity and noise resilience. Only with such an integrated theoretical framework will it become possible to navigate the vast chemical space available to molecular materials and identify those candidates that balance performance, stability, scalability and ease of fabrication.

\section{Multiscale methodology for rational design}
Designing organic memristive materials requires a methodological framework capable of spanning several orders of magnitude in spatial and temporal resolution. The behavior of a memristive molecule cannot be understood at a single level of theory: its electronic structure originates at the quantum scale, its adaptive response arises from collective ionic or conformational dynamics, and its characteristic hysteresis emerges at the macroscale, when these microscopic processes are integrated into a device-level environment where electric fields, temperature gradients and interfaces play an essential role. The complexity of this multiscale coupling is precisely what, up to now, has prevented the emergence of a coherent design roadmap for molecule-based memristors. A perspective that aspires to fill this gap must make explicit how theoretical tools can be articulated into a cumulative workflow where each level informs and constrains the next. Within this approach, one of the advantages offered by theoretical/computational Chemistry is to allow a significative methodological benchmarking. Specifically, in the critical point which is to clearly define how certain parameters relate and flow between the different levels of calculations, from the most detailed atomistic ones to Finite elements equations. In this aspect, to establish a correspondence between the macroscopic parameters obtained after the largest scale calculations  and the desired electronic memristive properties of these materials is the main goal. These parameters can therefore be used to sift through molecular libraries and to rationally choose new molecular proposals to be likely to present memristive behavior.

The starting point for all the cases is based on quantum atomistic calculations (Density Functional Theory, Ab-initio methods) where the electronic properties of the individual atoms conforming the molecular system are explicitly considered in the calculation (which requires considering $<500$ atoms in the best cases). In a first step Quantum-level calculations offer both detailed information about the electronic structure of single molecules and a methodological benchmarking which can be performed on simple representative molecules. This level is usually addressed with codes such as Gaussian,\cite{g16} (TRAN)SIESTA,\cite{Soler2002} ORCA\cite{neese2022} and MOLCAS.\cite{aquilante2016} In this step, nuclei and electrons are considered explicitly, the electronic interaction is treated for a certain distribution of nuclei following quantum mechanics either solving an approximated Schrodinger equation (wave-function methods) or Kohn-Sham equations, including time-dependent density functional theory (TD-DFT). Thus, obtaining the electronic properties of each individual molecule, its intramolecular bonding and any intermolecular interaction with the nearest neighbors is the objective at this level of study. This stage includes also the identification of the most suitable parameters including, exchange-correlation functional (EX-C), atomic basis size, adequacy of the periodic boundary conditions (PBC), size of the active space and relativistic corrections for spin orbit coupling consideration, among others. All these parameters should be selected to offer an accurate enough description of the molecular system at this level of detail so the rest of the procedure can be built upon, but also considering an acceptable computational cost. This initial set of calculations provides a general set of parameters such as: relaxed molecular structure, vibrational spectra (IR), magnitude of the intermolecular interactions, etc.  Also transport properties of single-molecule ensembles are presented at this calculation level, which is usually performed using Non-Equilibrium-Green-Functions (NEGF) methods. This latter calculation allows to extract directly the relationship between the molecular structure and the presence of one or various transmittance levels at the desired energies. 

\begin{figure}[ht]
 \centering
       \includegraphics[width=1\textwidth]{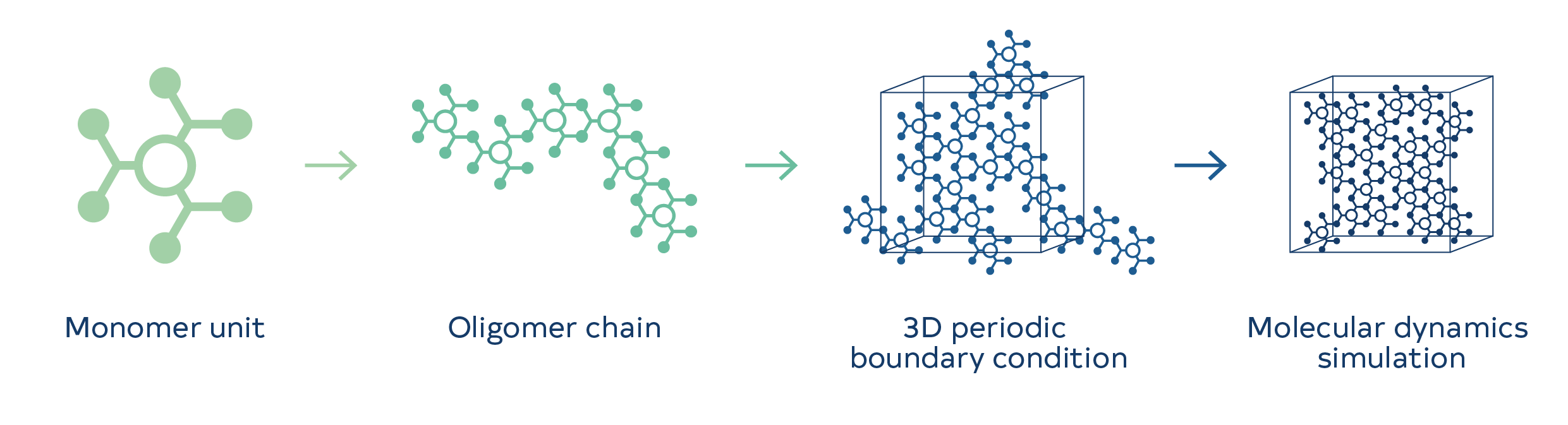}
\caption{Pictorial representation of a polymer MD simulation considering periodic boundary conditions.}
\label{fig2}
\end{figure}

The second step entails molecular dynamics (MD) simulations where the atomic trajectories for molecular systems made of up to $10^4-10^5$ particles (atoms) are obtained using the resolution of Newton’s motion equations. In this computational simulation, the molecular system is comprised of explicit atoms (nuclei coordinates), but the electron correlation (interaction of one electron with the rest) is ignored. MD methodology relies on molecular mechanics to estimate the potential energy of the molecular system. Most of the computational time is dedicated to obtaining the gradient of potential energy with respect to the coordinates of the atoms nuclei for a given configuration of the atoms. This calculation is performed for each time step on a MD simulation. MD programs make use of specific force fields that refer to the specific functional form and set of parameters used to compute the potential energy of a molecular system, and this applies to atomistic or coarse-grained systems. The contribution to potential energy for several variables can change depending on the force field. Most of them include bonded terms (interactions between atoms linked by covalent bonds) and non-bonded terms (long-range electrostatic and van der Waals interactions) where only interactions between atoms in different molecules, or between atoms in the same molecule separated by at least three bonds. (see figure \ref{fig2})

To choose force fields for each material it is crucial to rely on the previous results obtained from quantum procedures. Here, the molecular trajectories are obtained by solving numerically Newton’s equations of motion. Due to its more simplified description of the atoms, study size is enormously increased up to tens of thousands of atoms. An additional benefit of this calculation step is the possibility of including external factors, like finite temperature or an electric field and observe the structural and electronic distortion of the whole system. Importantly, at MD level, the temporal evolution of the system is considered for the first time, although the cost of the calculations makes prohibitive to go beyond a few microseconds with detail. The computational codes available for this stage are GROMACS\cite{Berendsen1984}, LAMMPS\cite{plimpton1995}, CP2K\cite{hutter2014} and AMBER.\cite{doi:10.1021/acs.jcim.3c01153} Additionally, the quantum effects on the results could be interrogated using ring polymer molecular dynamics (RPMD) simulations, which allows to include a statistical description of nuclear quantum effects in classical MD simulations.\cite{habershon2013} 
Combined Quantum Mechanics/Molecular Mechanics (QM/MM) methods, where selected atoms are set aside for the calculation of their electronic structure, and the rest of the system is treated with simpler molecular mechanics approaches, may also be convenient at this point (this approach is usually known as ONIOM-like methodology).\cite{chung2015}.    

\begin{figure}[ht]
 \centering
       \includegraphics[width=1\textwidth]{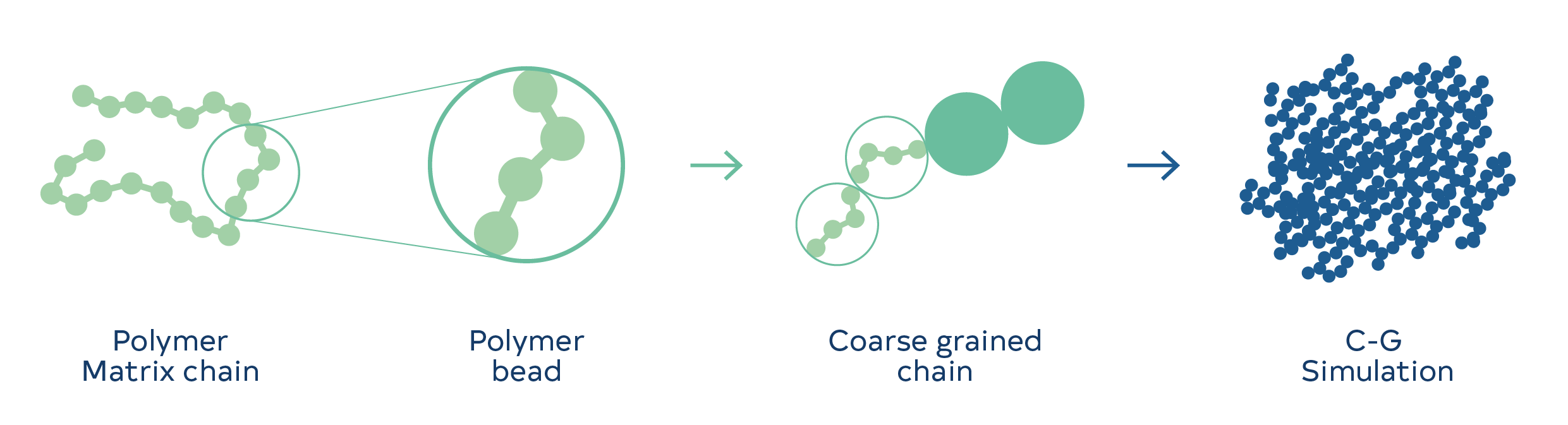}
\caption{Grain size determination and simplified polymer structure considering RIS model.}
\label{fig3}
\end{figure}

The next step consists in the so-called coarse-grained approach (CG), where the atoms are grouped into ‘pseudo- atoms’ where the fine structure is hidden, and the point of importance is the macromolecular accommodation to the application of external stimuli or to the presence of other ensembles. Coarse-grained molecular dynamics (CGMD) is the most popular technique in mesoscale modelling and allows to both increase the number of active participants in the memristive behavior and to introduce the time-scale with cheaper calculations (up to 10$^6$ atoms).\cite{christen2006, Joshi2021}.  This is a critical step, both for material chemistry and nanoelectronics, how to properly define the grains and which should be the explicit and implicit properties for them. In this case, the input will be well-based on the parameters that govern the fine interactions from the two previous calculation levels in order to properly use CGMD to simulate the mesoscale time-and-field dependent physics of the system. (see figure \ref{fig3}, Rotational Isomeric State (RIS) model) The computational codes that will be used at this step are ABCLUSTER,\cite{Zhang2015, Zhang2016} ESPResSo\cite{Weik2019} and GFN-xTB.\cite{doi:10.1021/acs.jctc.8b01176} Along with these, all the previous MD codes permit a ‘coarse-grained’ implementation so they will be also used at this calculation level. By adding this step to our methodology, one virtually has a complete memristive device modeled in a dynamic situation, with both the effect of electric field and temporal evolution being considered at the same time.

 To finally complement this multiscale methodology, effective and semiempirical differential equations based on finite elements/finite differences can be implemented in the procedure. This inclusion allows to predict the macroscopic electrical memristive behavior using the local parameters obtained from previous calculation steps. To finalize and complement the multiscale methodology, a mathematical model should be stablished based on the extracted parameters from the previous Ab-initio studies to model the working mechanism of the device under various operating conditions. An important advantage of these models is to efficiently implement semiempirical differential equations based on finite elements/ finite differences methodology. The finite elements method is a mathematical numerical approximation where the whole complex problem is divided into smaller parts. Then, these parts are modeled by simple equations that afterwards are assembled into a larger system of equations which considers the whole system. This allows to predict the macroscopic electrical memristive behavior by using the previous parameterization about the active region in the device and the molecular properties (obtained in the previous, more detailed, levels of calculation). In addition to these procedures, using the Monte Carlo method in the last step of the multiscale procedure is proposed. Monte Carlo methodology is based on random numerical sampling to solve deterministic problems that can be represented by a probabilistic interpretation (in our present case, probability distribution of material structures under external perturbances).\cite{Metropolis1949} With it, the device active region can be simulated by using mean field approximation and by letting the particles to interact in a manner that is consistent with the previous parameters. A set of possible microstates forming a stationary probability distribution is found which, following the ergodic theorem, represent the result of an empirical measurement on a random state of the system. Thus, macroscopic magnitudes of interest (such as clustering ability and intermolecular interaction) of the memristive device could be determined.\cite{gutierrez2025copas}
To finalize the procedure various critical features of memristive materials should be revised within the calculations prediction: Number of accessible conductance states, cycling endurance, energy consumption, reproducibility, chemical stability, resistance on/off ratio and expected switching time.

The three types of materials proposed (Ionic-, Redox-and Magnetic-based materials) will be tentatively assessed based on these capacities. With this we will analyze and classify these materials following a criterion of ‘adequacy for memristive applications’.  In this point we will be able to answer a decisive question: how to rationally build molecular memristors with enhanced properties by using chemical design. The knowledge gained by identifying which properties are linked to better outcomes in memristive behavior will permit us to employ a strategy of High-Throughput Virtual Screening (HTVS). This computational tool has been proven extremely useful in pharmaceutical science for drug design and is based on an automatic search of molecules/materials in the chemical space. HTVS avoids countless trial-and-error experiments by carrying out a systematic exploration of the molecular space for leads. The procedure entails performing a crib through a finite subsection of space (in our case of molecular materials) in order to select molecular candidates that comply with the selected descriptors so that they are more likely to have the optimized properties distinctive of materials with good memristive capabilities. In this funnel-like approach, an increasingly strict criterion removes molecules that are not of interest and keeps the highest performing candidates. 

With the previous information it is possible to extend all the knowledge that has been built throughout the approach into a useful computational tool to perform an automatic virtual screening of memristive materials. The goal is to elucidate and quantify the underlying relationships between structural/composition and physical properties obtained from large databases containing the theoretical properties of existing and hypothetical materials. The best-known methods for this data-driven materials search are two: the former includes the selection of the appropriate QSPR (quantitative structure-property relationship) descriptors governing, in our case, the memristive behavior. After that, an objective function is defined based on linear correlations of a set of variables of the material, maximizing the desired behavior. An optimization solution is applied to the initial large data set of materials being analyzed and finally the objective function is applied to select the best candidates. The second method incorporates the newest machine learning techniques applied to fast screening and materials discovery. Here, genetic algorithms are applied, consisting in metaheuristic optimization inspired by Darwinian evolution. Performing crossover, mutation and selection operations, this algorithm produces a population of evolving candidate solutions. For this the scientific community makes use of materials libraries online such as AFLOW (http://aflowlib.org), Materials Cloud (https://www.materialscloud.org/home), Materials Project (https://materialsproject.org), NoMAD Repository (http://nomad-repository.eu) , Open Quantum Materials Database (http://oqmd.org) and Computational Materials Repository (https://cmr.fysik.dtu.dk), where the electronic and physical properties of organic (and inorganic) materials are tabulated. The main idea is to have a visual interface in place where we will explore the chemical diversity of certain families of complexes and find the most suitable example that optimizes one or more particular characteristics. 

Once the pathway through the general method is clear, it is important to particularize to each of the three mentioned mechanisms: Ionic migration, redox-activity and interplay between magnetism and electronic current. Within each mechanism we have chosen a representative organic material related with the most successful experiments found in the literature.

\subsection{Specific approaches for mechanisms based on ionic migration}

The author's proposal to obtain ionic migration–based memristive materials use a composite formed by a molecular ionic salt with a conductive polymer, and optionally includes an additional polymer to facilitate internal ion migration. When a voltage is applied, mobile ions are displaced toward the electrode interfaces, thereby decreasing the injection barrier and modulating the current through the polymer matrix as a function of both time and bias, with a remanent response characteristic of memristive behavior. This phenomenon is analogous to vacancy-driven effects in inorganic systems such as nonstoichiometric TiO$_{2-x}$, and its rational design requires establishing clear structure–property relationships. To this end, high-level quantum-chemical calculations (DFT and wave-function methods) on specific cation/anion pairs are proposed to quantify intermolecular interactions, preferred geometries, and atomic charge distributions. Systematic chemical modification of the imidazolium cation, varying alkyl chain length and introducing electron-donating, electron-withdrawing, or $\pi-\pi$ stacking substituents, is used to tune electrostatic interactions, aggregation propensity, and cluster formation within the polymer matrix, which are then further investigated using periodic simulations with and without external electric fields to study ion separation, cluster stability, and field-induced reorganization. Subsequently, solid-state DFT–NEGF calculations on ion-decorated metallic electrodes (e.g., Au, Ag, Al) embedded in the polymer can be employed to evaluate adsorption, structural relaxation, work-function shifts, and transport properties as functions of coverage, adsorption geometry, and ion chemistry, using a tailored exchange–correlation strategy to achieve structural convergence between the metallic surface and the molecular ions.


In parallel, a time-dependent numerical framework for charge transport in ionic-liquid-based memristive devices is proposed to extend existing steady-state models toward dynamic operation. The first modeling route involves a self-consistent solution of Poisson’s equation coupled to carrier transport equations (e.g., drift–diffusion formulations derived from the Boltzmann transport equation), explicitly incorporating structural disorder and intermolecular interactions relevant to organic systems such as light-emitting electrochemical cells. A complementary kinetic Monte Carlo approach is employed to track the stochastic evolution of ion positions and energies under applied electric fields across different time scales, including elastic and inelastic scattering processes parametrized from the preceding quantum-chemical and solid-state studies.\cite{gutierrez2025copas} The resulting spatiotemporal ion distributions and current–voltage responses are then used to derive an equivalent circuit representation of the device, suitable for implementation in standard circuit simulators (e.g., SPICE). Finally, this circuit-level description is expanded to design and analyze a 3-to-1 perceptron-like building block compatible with CMOS integration, enabling the use of ionic memristive elements in neuromorphic architectures.

\subsection{Specific approaches for mechanisms based on redox switching}

Herein, multistate redox-switching behavior in molecular materials will be investigated using single-metal redox-active complexes of the archetypal \ce{[TM(phen)_3]^2+} family, where TM = \ce{Fe^2+}, \ce{Ru^2+}, or \ce{Ir^3+} and phen is 1,10-phenanthroline. The number of accessible redox states will be enhanced by replacing one or more phenanthroline ligands (redox-inert) with redox-active phenylazo-pyridine units, and by systematically probing the effect of different counterions on redox-memory retention, including steric tuning via increasing size counteranions.  These systems are expected to display multistate switching with a large number of discrete conduction levels stabilized by distinct counterion geometries, where anion reorganization induces barriers to redox reversibility and hysteresis in the current–voltage response. 
Molecular electronic structure and redox potentials were correlated with the number of accessible conducting states by means of TD-DFT calculations, varying the nature of the transition-metal center, the degree of azo substitution, and the size of the counterion. In addition, the (phenylazo)pyridine ligand was further modulated with electron-donating and -withdrawing substituents to identify combinations of metal ion, ligand set, and counterion that maximize the density of distinct conducting states within the relevant energy window. Intermolecular interactions and counterion distributions in extended assemblies were then investigated under an external electric field using a multilayer ONIOM-type strategy in order to assess how these supramolecular features, influence bulk redox properties and memristive performance and to refine the molecular design.
The role of vibrational structure and Joule heating in cooperative redox processes at the macroscopic scale was examined by explicitly coupling electronic transitions to both molecular and lattice vibrational modes. Single-molecule vibrational spectra for different counterion arrangements were benchmarked against periodic simulations of the molecular crystal, enabling the construction of an effective vibronic model and complementary coarse-grained molecular dynamics analyses of field-induced vibro–electronic coupling in electrode-confined systems. Finally, the molecular orbitals obtained in the preceding steps were related to the density of states and transmission characteristics of complexes adsorbed on Au, Ag, and Al electrodes by means of geometry-relaxed DFT–NEGF calculations that explicitly account for counterion positions and their evolution under bias. Structural relaxation at each applied voltage, with and without counterions, allowed the underlying structure–transport relationships in redox-based molecular memristors to be established and yielded design principles for new materials exhibiting a high number of stable, addressable conduction states at low voltages.

\subsection{Specific approaches for mechanisms based on helical magnetic molecules}

The interplay between electronic conduction and magnetism in helical molecular systems also offers a promising platform for achieving memristive behavior in molecular materials. In these systems, chirality-induced spin selectivity (CISS) enables significant spin polarization of electrons in the absence of traditional magnets. Lanthanoid-binding-tag (LBT) metallopeptides serve as ideal model compounds to explore this effect, as they form helical structures capable of coordinating various transition metals and lanthanoids. Spin-polarized currents passing through these helical molecules interact with the magnetic moment of the coordinated ion, resulting in incomplete spin relaxation under rapid voltage switching. This memory of spin orientation manifests as a current dependent on the voltage history, fully consistent with memristive behavior. Detailed studies on \ce{Tb^3+}-complexed LBT have been key to understanding how the molecular helicity governs the magnetic polarization of the embedded ion, laying the foundation for magnetic memristor design.\cite{Cardona-Serra2021}

At the electronic and structural level, density functional theory (DFT) and time-dependent DFT (TD-DFT) calculations may reveal the magnetic orbital spin splitting induced by the interaction between the metal ion’s spin and the chiral ligand field. Calculations on helical polypeptides complexed with diamagnetic \ce{Y^3+}, high-spin \ce{Mn^2+}, and highly anisotropic \ce{Dy^3+} further elucidate how metal ion identity influences magnetic exchange and electron polarization within the helix. In parallel several other investigations are currently exploring how the chemisorption of diamagnetic helical peptides on metal electrodes affects surface electronic structure and induces spin polarization. Such studies emphasize the relationship between helical length and spin filtering efficiency, suggesting that longer helices, such as those derived from large polypeptides, enhance magnetic polarization and memristive functionality. In other hand, performing Non-equilibrium Green’s function NEGF–DFT transport calculations on metallopeptide junctions may connect electronic structure to transport signatures, confirming the role of magnetic coupling and chirality-induced spin selectivity in controlling spin-dependent electron transmission.

At the macroscopic scale, spin relaxation dynamics in these helical systems are modeled via effective time-dependent differential equations incorporating electromagnetic interactions. Finite-element simulations parameterized by electronic structure and transport results enable predictions of how molecular spin states evolve under applied polarized currents and how these dynamics modulate device current-voltage characteristics. This multiscale approach rationalizes the coupling between molecular magnetic exchange interactions and macroscopic spin-filtering effects responsible for memristive response. Together, these findings highlight the potential of helical magnetic molecules as building blocks for next-generation spintronic memristors, offering a pathway to devices with intrinsic, voltage-history-dependent spin-polarized conduction.

\section{Perspectives}

The landscape of organic memristive materials is still in its formative stage, and this Perspective highlights both the remarkable opportunities and the conceptual gaps that define the field. Organic systems offer an unparalleled degree of chemical freedom, ion mobility tuning, redox molecular engineering, and spin–chirality interplay, that can be harnessed to create neuromorphic elements far more adaptive, energy-efficient and molecularly tailored than current inorganic technologies. Yet capitalizing on this potential requires a shift from exploratory experimentation toward a design philosophy grounded in predictive theory.

The three mechanisms discussed—ionic migration, redox-active switching and magnetic–chiral conduction—illustrate the breadth of physical phenomena available to organic systems, but also reveal a shared challenge: memristive behaviour is inherently multiscale. It emerges from quantum-level electronic structure, mesoscale ionic or structural dynamics, and macroscopic device operation under realistic fields and temperatures. A future roadmap for the field must therefore embrace theoretical integration rather than isolated modelling efforts.

The multiscale methodology proposed here aims to serve precisely this purpose. By connecting quantum chemistry, molecular dynamics, coarse-grained simulations and device-level finite-element modelling, it becomes possible to construct a coherent pipeline capable of translating molecular design into functional device predictions. This framework is useful to rationalize existing experimental results while also sets the stage for systematic exploration of the vast chemical space available to organic materials.

Looking ahead, several opportunities emerge. First, the development of several descriptors, ion–polymer tunable interaction, redox-state electronic densities, magnetic relaxation regimes, interfacial work-function modulation, will be crucial for enabling meaningful comparison across different material families. Second, integrating these descriptors into high-throughput virtual screening strategies could accelerate the discovery of optimized candidates with tailored synaptic functions such as multilevel memory, controlled volatility or ultra-low switching energy. Finally, the convergence between cheminformatics, machine learning and multiscale simulation may ultimately give rise to a true “materials-by-design” paradigm for neuromorphic chemistry.

Organic memristors are poised to become a foundational platform for next-generation neuroinspired hardware, but their maturation will depend on our ability to unify chemical intuition with rigorous theoretical design. By articulating the key mechanisms and outlining a computational roadmap, this Perspective seeks to contribute to that transition—from promising demonstrations toward truly programmable, molecularly engineered neuromorphic materials.

\ack{S. C.-S. acknowledges Dr. Alejandro Gaita-Ariño for his stimulating conversations, continuous encouragement, and unwavering support for exploring new ideas, which has been a constant source of motivation throughout this work.}
 




\bibliographystyle{vancouver} 
\bibliography{refs} 

\end{document}